\documentclass[epj]{webofc}
\usepackage[utf8]{inputenc}
\usepackage[varg]{txfonts}   
\usepackage{booktabs}
\usepackage{xcolor}
\definecolor{darkred}{rgb}{0.4,0.0,0.0}
\definecolor{darkgreen}{rgb}{0.0,0.4,0.0}
\definecolor{darkblue}{rgb}{0.0,0.0,0.4}
\usepackage[bookmarks,linktocpage,colorlinks,
    linkcolor = darkred,
    urlcolor  = darkblue,
    citecolor = darkgreen]{hyperref}
\usepackage{subfigure}
\wocname{EPJ Web of Conferences}
\woctitle{Lattice2017}
\begin{document}
%
\selectlanguage{english}
\title{%
Topological Susceptibility in $N_f=2$ QCD at Finite Temperature
}
\author{%
\firstname{Sinya} \lastname{Aoki}\inst{1} \and
\firstname{Yasumichi} \lastname{Aoki}\inst{2,3} \fnsep\thanks{Speaker, \email{yasumichi.aoki@kek.jp}}\and
\firstname{Guido}  \lastname{Cossu}\inst{4} \and
\firstname{Hidenori}  \lastname{Fukaya}\inst{5} \and
\firstname{Shoji}  \lastname{Hashimoto}\inst{2,6} \and
\firstname{Kei}~~\lastname{Suzuki}\inst{2} 
\hspace{6pt}
(JLQCD Collaboration)
}
\institute{%
Yukawa Institute for Theoretical Physics, Kyoto University, Kyoto
606-8502, Japan
\and
High Energy Accelerator Research Organization (KEK), Tsukuba, Ibaraki 305-0801, Japan
\and
RIKEN BNL Research Center, Brookhaven National Laboratory, Upton, NY
11973, USA
\and
School of Physics and Astronomy, The University of Edinburgh,
Edinburgh EH9 3JZ, United Kingdom
\and
Department of Physics, Osaka University, Osaka 560-0043, Japan
\and
SOKENDAI (The Graduate University for Advanced Studies), Tsukuba, Ibaraki
305-0801, Japan
}
\abstract{%
We study the topological charge in $N_f=2$ QCD at finite temperature
using M\"obius domain-wall fermions.
The susceptibility $\chi_t$ of the topological charge defined
either by the index of overlap Dirac operator or a gluonic operator
is investigated at several values of temperature $T\ (>T_c)$ varying
the quark mass. A strong suppression of the susceptibility is
observed below a certain value of the quark mass.
The relation with the restoration of $U_A(1)$ is discussed.
}
\begin{flushright}
 KEK-CP-364, RBRC-1257
\vspace{-12pt}
\end{flushright}
\maketitle
\section{Introduction}\label{sec:intro}

Topological susceptibility in QCD at finite temperature has acquired much  
attention recently due to its phenomenological interest.
Mass of the QCD axion, one of the candidates of dark matter, is given by
the topological susceptibility, and its dependence on temperature 
determines the abundance of the axion in the universe.
A quantitative estimate can in principle be provided by lattice QCD,
and was one of the topics of the panel discussion of this
year's lattice conference
\cite{Moore:2017ond,Bonati:2017nhe,Lat2017KovacsPanel,Lat2017FukayaPanel}.
This study is not meant to provide some quantitative results at
phenomenologically important temperatures $500\lesssim T\lesssim 1000$
MeV \cite{Moore:2017ond},
but rather to understand the nature of the phase transition 
in two-flavor QCD \cite{Lat2017FukayaPanel}.

The fate of the $U_A(1)$ symmetry at and above the phase transition
for vanishing $u$ and $d$ quark masses is one of the long standing and
fundamental questions in QCD.  While at any temperature the $U_A(1)$ chiral
anomaly exists, manifestation of the $U_A(1)$ breaking is only possible
if the gauge field configurations with non-trivial topology actually
have non-vanishing contribution. The non-trivial QCD
configurations also produce the topological susceptibility. Thus there is
naturally a link in between these two physical quantities.

One powerful theoretical approach for these problems is to use the
properties of the spectrum of the Dirac operator
\cite{Cohen:1996ng,Cohen:1997hz,Lee:1996zy,Evans:1996wf}.
Along this line Aoki, Fukaya and Taniguchi (AFT) revisited the problem
assuming the overlap fermions for the UV regulator for quarks 
\cite{Aoki:2012yj}.
They claim that the $U_A(1)$ symmetry in two flavor ($N_f=2$) QCD is
recovered in the chiral limit for  
temperatures at and above the critical one. Furthermore, the derivatives of
the topological susceptibility with respect to the quark mass $m$ vanish
at any order. 
It means
that the susceptibility, 
which is zero at the chiral limit, stays zero in the vicinity of $m=0$.
As the susceptibility is non-zero for infinitely heavy quarks,
there must be a critical mass which divides the regions with zero and
non-zero topological susceptibility.

The relation of the spectrum of the Dirac operator with $U_A(1)$ was also
studied by
Kanazawa and Yamamoto (KY) more recently \cite{Kanazawa:2015xna}.
Assuming the $U_A(1)$ breaking they derived a relation between the
$U_A(1)$ susceptibility, which is a measure of the $U_A(1)$ breaking, and
the topological susceptibility through a low energy constant. 
According to their study, the
topological susceptibility should be proportional to the squared quark
mass, thus, should exhibit quite different mass dependence to that of
AFT. Kanazawa-Yamamoto claims the assumption that the
spectral density is analytic near the origin in AFT needs
to be abandoned to have the $U_A(1)$ breaking. The analyticity, however,
seems intact in the simulations with overlap fermions \cite{Cossu:2013uua}
and domain wall fermions with overlap-reweighting
\cite{Cossu:2015kfa,Tomiya:2014mma,Lat2017Suzuki}, 
which have exact chiral symmetry. 

Studying the topological susceptibility in depth would
add another dimension for the understanding of the nature of the finite
temperature transition in $N_f=2$ QCD, especially if it is done in
conjunction with the direct measurement of the $U_A(1)$ breaking.
Also, understanding the fate of the $U_A(1)$ breaking should be important 
for the computation of the topological susceptibility to a required
precision necessary for phenomenology.

Chiral symmetry plays a crucial role for the study of $U_A(1)$
\cite{Cossu:2015kfa,Tomiya:2014mma}. We use the M\"obius domain
wall fermion and reweighting method to the overlap fermion ensemble.
In this report and the one for the $U_A(1)$ breaking
\cite{Lat2017Suzuki}, the main lattice spacing used is finer than we have
used in \cite{Cossu:2015kfa,Tomiya:2014mma}. This helps to reduce the
residual chiral symmetry breaking of the domain wall fermions and to
make the reweighting efficient.

The AFT scenario suggests a critical mass $m_c>0$ which divides the regions
of topological charge zero and non-zero. If this is true it is
consistent with the first-order phase transition
\cite{Aoki:2012yj,Aoki:2013zfa}, which was suggested by
Pisarski and Wilczek \cite{Pisarski:1984ms} for the case of the $U_A(1)$
restoration. 
This could, then, change the widely believed phase diagram, called
the Columbia plot at the upper-left corner. If similar dynamics exists at the
physical strange quark mass point, it would affect the nature of the transition
of the physical point depending on the value of $m_c$.




This report is organized as follows. In Sec.~\ref{sec:method}, 
the calculation set-up and methods are described. Starting with a discussion
on the sampling of the topological charge, an elaborate estimate of the error
for the topological susceptibility is explained in Sec.~\ref{sec:results},
followed by our main results. Sec.~\ref{sec:summary} is devoted to
summary and outlook. We use $a=1$ units throughout. 
All the results reported here are preliminary.

\section{Methods and parameters}\label{sec:method}

Our simulation is carried out using M\"obius domain wall fermions 
for two dynamical quark flavors \cite{Cossu:2015kfa}.
A particular focus is placed on $N_t=12$, $\beta=4.3$ ensembles with 
five different masses in this report. The corresponding temperature is
$T\simeq 220$ MeV. At a fixed $\beta$ value, two different temperatures
$N_t=8$ and 10 are examined and results are reported. As a check of finite
lattice spacing effects, a coarser lattice at $\beta=4.1$ and $N_t=8$
corresponding to $T\simeq 220$ MeV is examined.
For all lattices reported here the spatial site number is $L=32$.

The lattice cutoff as a function of $\beta$ for  these lattices is
obtained with the Wilson flow scale $t_0$ using the zero temperature results
and an interpolation \cite{Tomiya:2016jwr}.

Topological susceptibility is defined as
\begin{equation}
 \chi_t = \frac{1}{V}\langle Q_t^2\rangle,
  \label{eq:chi_t}
\end{equation}
where $V$ is the four dimensional volume and $Q_t$ is the topological charge.

We examine two definitions of the topological charge.
One is the space-time sum of the gluonic topological charge density 
after the Symanzik flow at $t=5$. The other is the index
of the overlap-Dirac operator \cite{Tomiya:2016jwr}.

As pointed out in \cite{Cossu:2015kfa,Tomiya:2016jwr},
it is essential to reweight to overlap ensemble from domain wall
\begin{equation}
 \langle {\mathcal O}\rangle_{OV} = 
  \frac{\langle {\mathcal O}R\rangle_{DW}}{\langle R\rangle_{DW}},
  \label{eq:reweighting}
\end{equation}
where $R$ is the reweighting factor defined on each gauge field
configuration, to correctly take into account the 
effect of (near) zero modes of the overlap-Dirac operator. Partial quenching
by the use of valence overlap operators on dynamical domain wall ensembles 
leads to an artificial enhancement of low modes. 
The topological charge defined through the zero mode counting suffers
from such artificial effects, which can be eliminated by the reweighting.

We investigate two definition of the topological charge on the
original domain wall ensemble and on the overlap ensemble generated
through the reweighting. Altogether,
four values of topological susceptibility are obtained at each
parameter point as shown in the next section.

We are aiming to acquire the data from 30,000 molecular dynamics time
units with hybrid Monte-Carlo simulation for each ensemble.  Some of the
reported data here are still undergoing improvement of statistics.

\section{Results}
\label{sec:results}

\subsection{Topological charge sampling and error estimate}
\begin{figure}[thb] 
 \centering
 \includegraphics[width=7cm,clip]{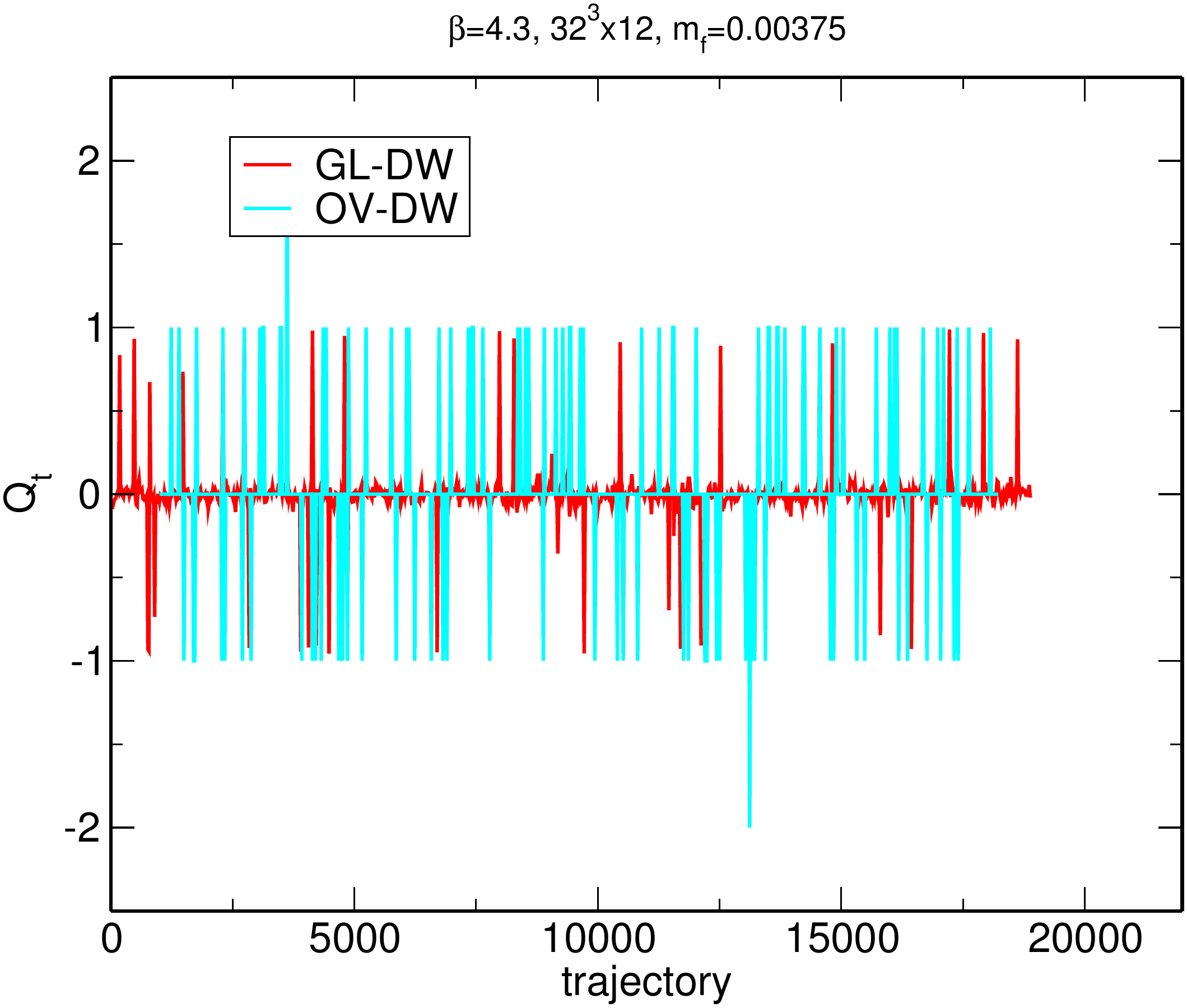}
 \includegraphics[width=7cm,clip]{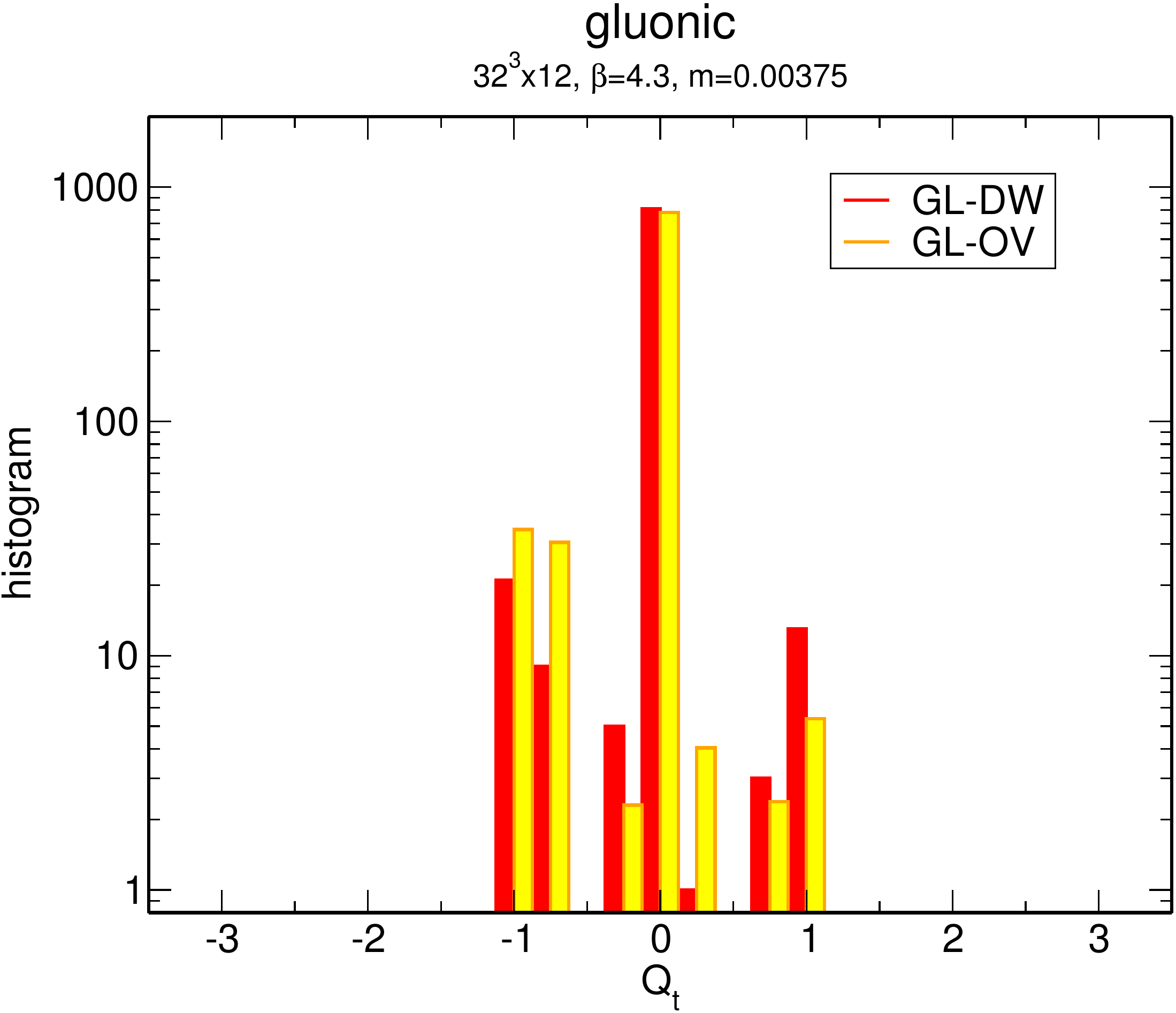}
 \caption{Monte-Carlo time history of topological charge (left) and
 histogram for gluonic measurement at $m=0.00375$ ($\simeq 10$ MeV) (right).}
 \label{fig:history}
\end{figure}

\begin{figure}[thb] 
 \centering
 \subfigure[$m=0.00375$ ($\simeq 10$ MeV). 
 \label{fig:hstg_OV_00375}
 ]%
 {
 \includegraphics[width=0.475\textwidth,clip]
 {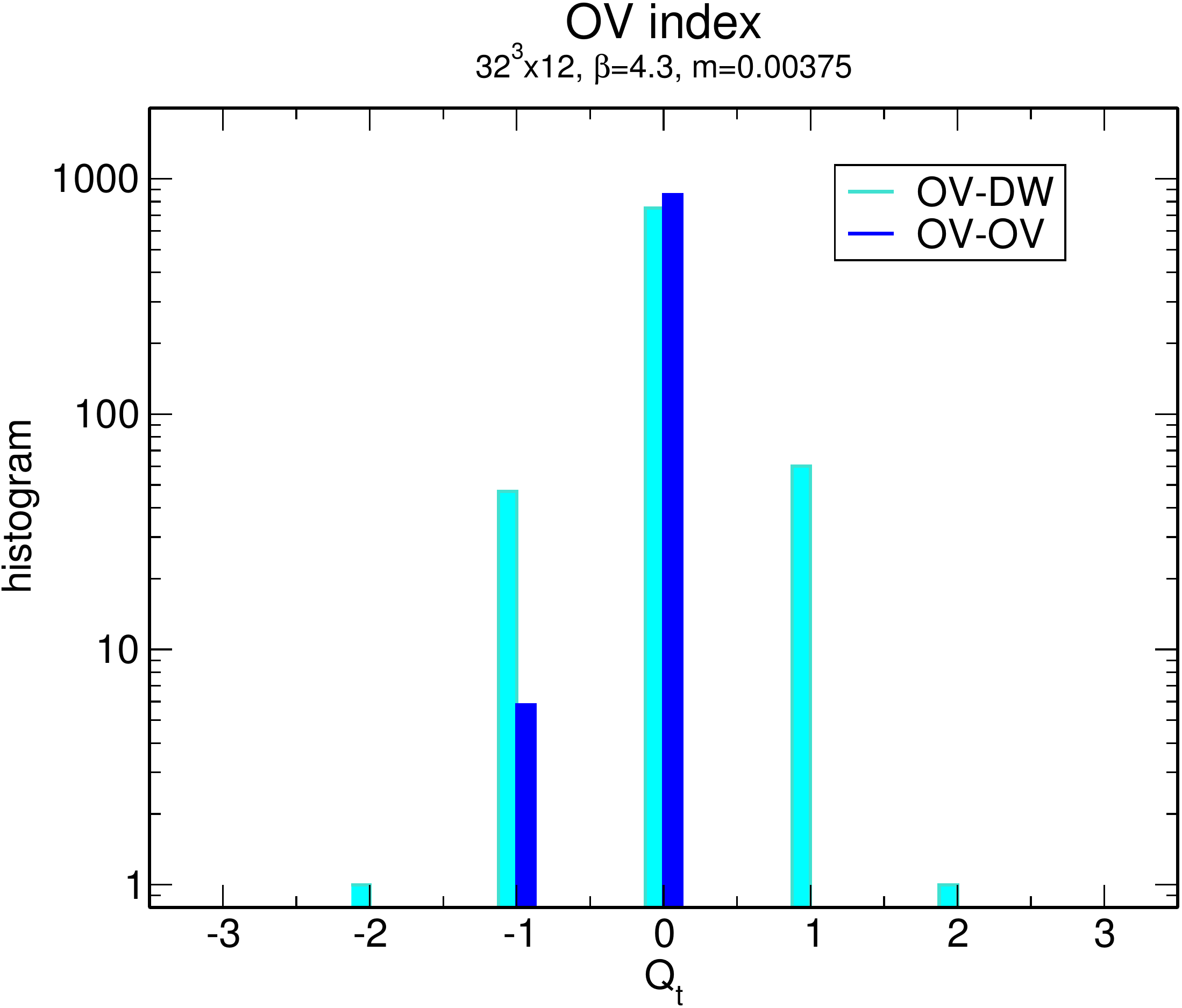}
 }
 \hfill
 \subfigure[$m=0.001$ ($\simeq 2.5$ MeV).
 \label{fig:hstg_OV_001}
 ]%
 {
 \includegraphics[width=0.475\textwidth,clip]
 {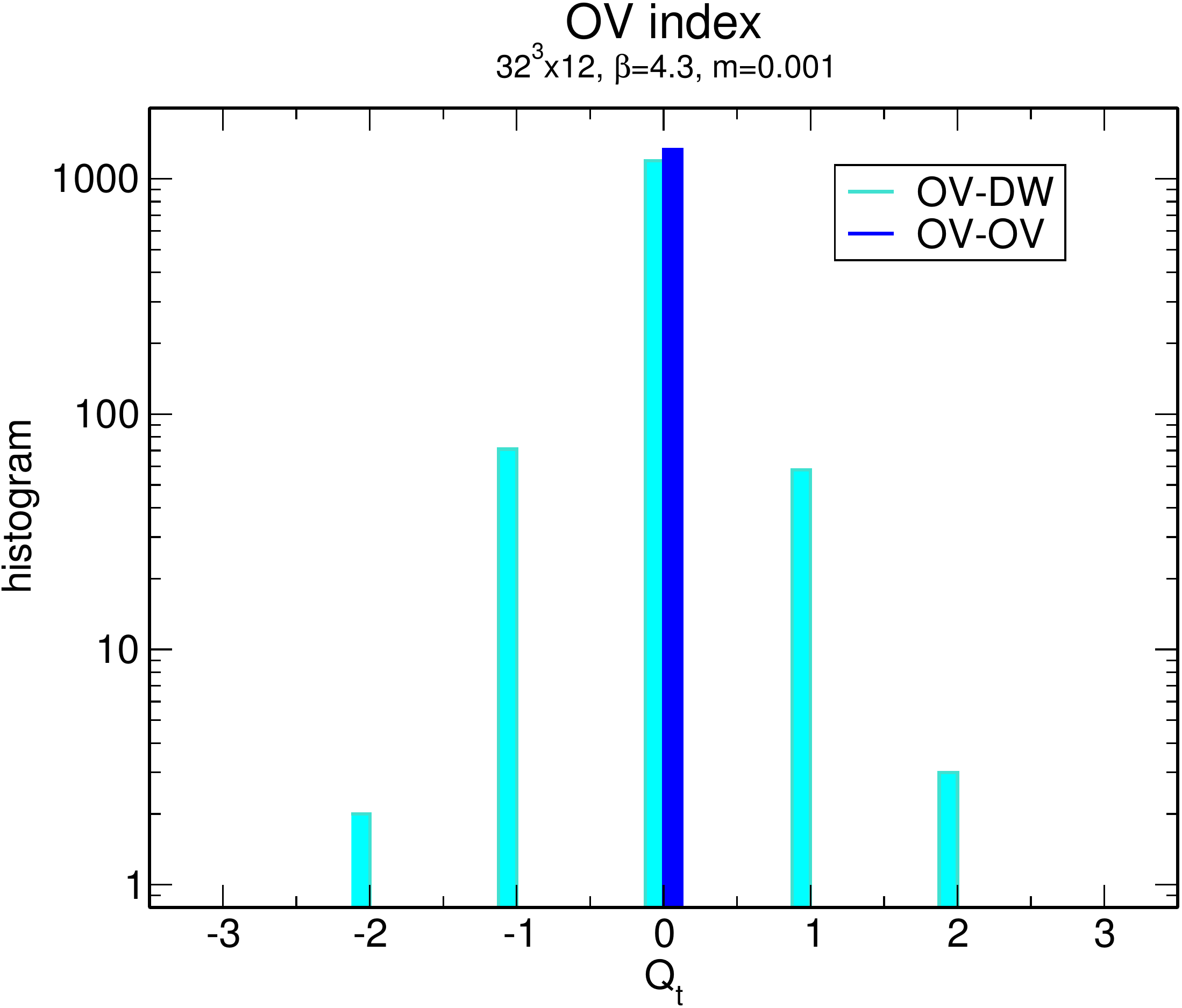}
 }
 \caption{Histogram of topological charge measured by the overlap index
 before (OV-DW) and after (OV-OV) the reweighting to overlap ensemble.}
 \label{fig:hstg_OV}
\end{figure}

The left panel of Figure \ref{fig:history} shows the Monte-Carlo time
history of the topological charge for $\beta=4.3$ with $N_t=12$ 
($T\simeq 220$ MeV) and bare mass $m=0.00375$ ($\simeq 10$ MeV) sampled
every 20th trajectory. 
One trajectory amounts to a unit time molecular dynamics evolution followed
by an accept-reject step. The red line corresponds to the charge
measured with the gluonic definition (``GL''), while cyan represents 
that with the overlap index (``OV''). The legends also show the ensemble
on which the calculations are based, which are domain 
wall (``DW'') for both. The right panel plots the histogram of the charge
from ``GL-DW'' and that after the reweighting to the overlap ensemble ``GL-OV''.
The bin size used can be read from the combined size of a pair of
neighboring red and yellow bars. It shows there is not much
difference between the data before and after the reweighting. 
Figure \ref{fig:hstg_OV_00375} shows the histogram of the topological
charge measured through the overlap index before (OV-DW) and
after (OV-OV) the reweighting. Here the width of the
distribution shrinks significantly after the reweighting.
This is due to the fact that the spurious zero modes on the domain wall
ensemble gets suppressed. On the other hand,
since such spurious zero modes are also suppressed by gauge field smearing,
there appeared less difference between the gluonic measurements before
and after the reweighting. From these data we calculate the topological
susceptibility from Eq.~(\ref{eq:chi_t}).

Special attention is required when there is no weight for the
non-trivial topology, shown in Fig.~\ref{fig:hstg_OV_001} as an example.
The OV-OV histogram shows that all samples fall in the $Q_t=0$ sector.
There actually is a non-zero $|Q_t|=1$ sample, but far
smaller than the minimum of the $y$ axis shown because of the small
reweighting factor. As a result, the topological
susceptibility is consistent with zero, with a jackknife
error $\chi_t= 4.4(4.4)\times 10^{2}$ MeV$^4$.
One should not take this as the sign of exact zero of $\chi_t$.
This situation is similar to null measurements of rare processes in
experiment. We estimate the upper bound of $\langle Q_t^2\rangle$
by imposing the condition that one measurement out of the full sample
had $|Q_t|=1$ value. 
If the number of samples is $N$, then the upper bound of the topological
susceptibility is
\begin{equation}
 \Delta'\chi_t = \frac{1}{N}\frac{1}{V}.
\end{equation}
With a reweighting, the effective number of samples gets reduced.
We use the following quantity for the number of samples after reweighting:
\begin{equation}
 N^{\rm eff} = \frac{\langle R\rangle_{DW}}{R_{max}},
\end{equation}
where $R_{max}$ is the maximum value of the reweighting factor in the
ensemble \cite{Tomiya:2016jwr}. 
As $\Delta'\chi_t$ can also be regarded as a resolution of the topological
susceptibility given the number of samples -- even if countable $|Q|>0$
sector exists as in the Fig.~\ref{fig:hstg_OV_00375} -- we estimate the 
corrected statistical error of $\chi_t$ for all the cases as
\begin{equation}
 \Delta\chi_t = \max(\Delta^{JK}\chi_t,\Delta'\chi_t),
\end{equation}
where $\Delta^{JK}\chi_t$ is the jackknife error of $\chi_t$.
For the case of Fig.~\ref{fig:hstg_OV_001}, $N^{\rm eff}=32$ out of a total
of 1326 samples measured every 20th trajectory. Now the error after this
correction reads $\Delta\chi_t=3.9\times 10^6$ MeV$^4$.

\subsection{Topological susceptibility at $T\simeq 220$ MeV}

\begin{figure}[thb]
 \centering
 \includegraphics[width=0.5\textwidth,clip]{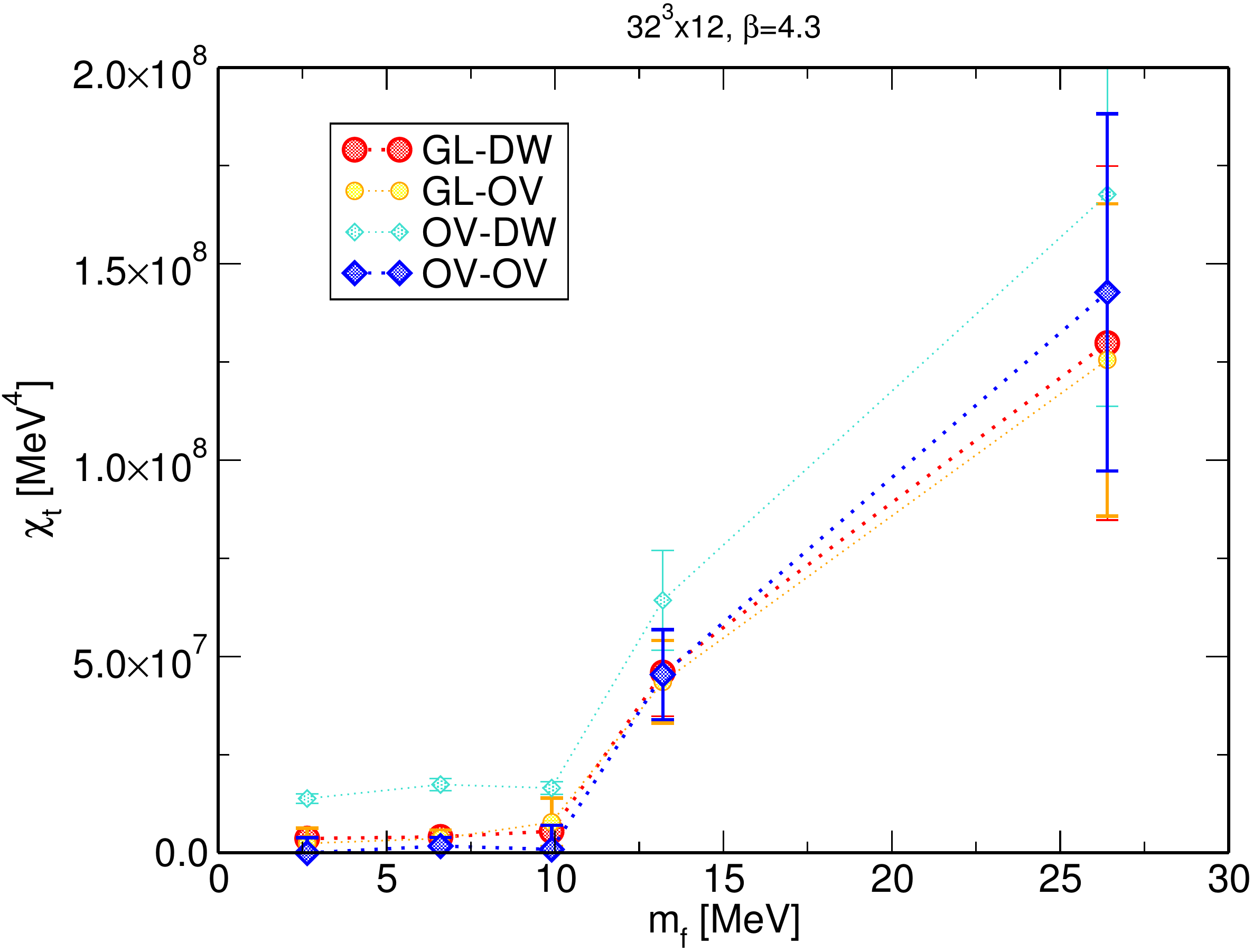}
 \includegraphics[width=0.45\textwidth,clip]{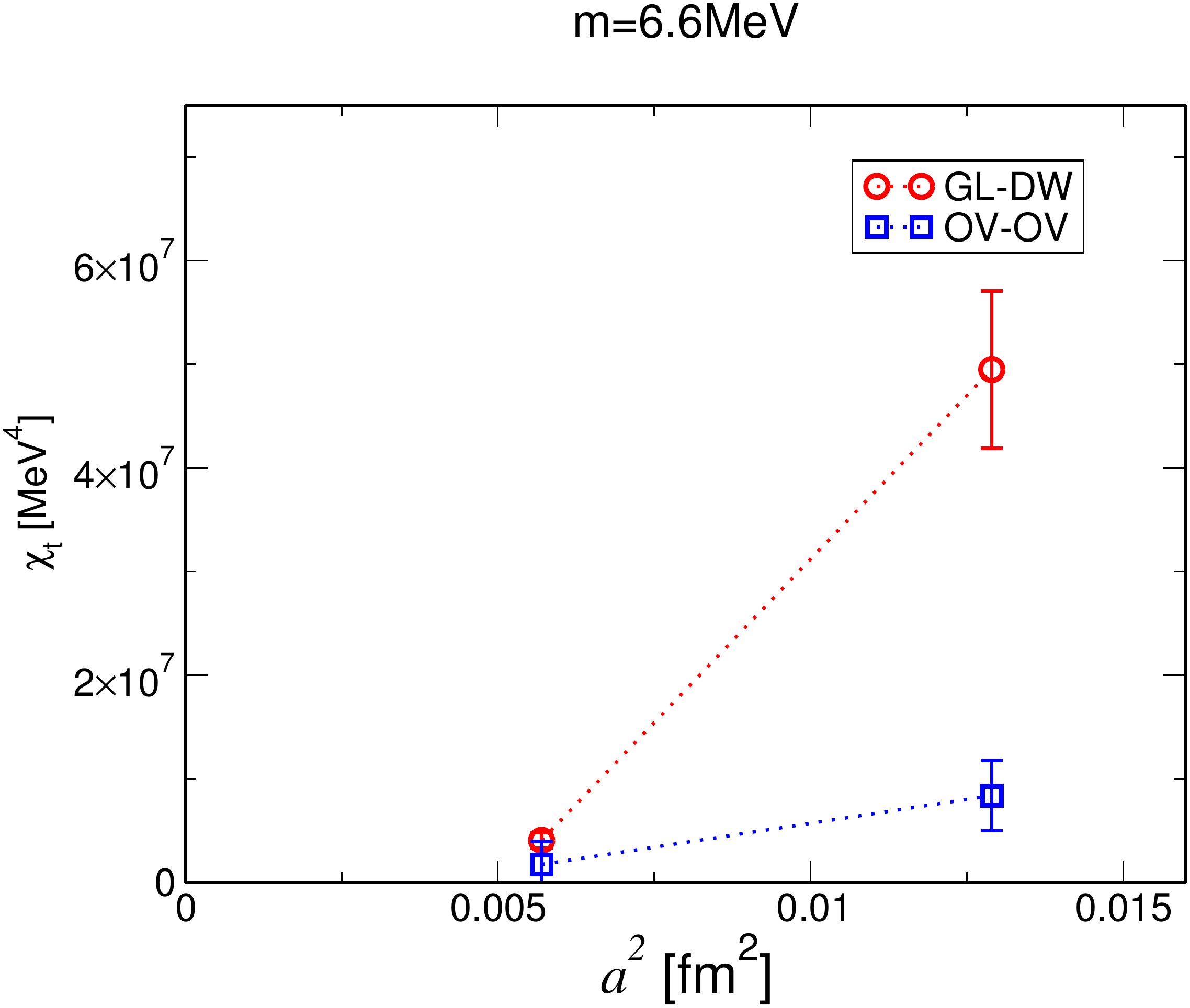}
 \caption{Topological susceptibility $\chi_t$ at $T\simeq 220$ MeV as
 function of quark mass (left) and $a^2$ dependence of $\chi_t$ at
 $m=6.6$ MeV ($ma=0.00375$ for finer lattice) (right).}
 \label{fig:chit220}
\end{figure}

The left panel of Fig.~\ref{fig:chit220} shows the quark mass dependence of
topological susceptibility for $N_t=12$ with $T\simeq 220$ MeV.
The color coding used here is the same as in the history and histogram shown
in Figs.~\ref{fig:history} and \ref{fig:hstg_OV}. As noted in the previous
section, 
OV-DW can yield enhanced fictitious zero-modes. Indeed, the cyan points
appear as outliers and the resulting $\chi_t$ gets fictitious enhancements.
Also, as mentioned for $m\simeq 10$ MeV, the histograms of GL-DW
and GL-OV are similar. Because of this, $\chi_t$ for GL-DW and GL-OV
appear consistent. As the reweighting reduces the effective number of
statistics, we use GL-DW in comparison with GL-OV.

The right panel of Fig.~\ref{fig:chit220} shows $\chi_t$ at $m\simeq 6.6$
MeV and $T\simeq 220$ MeV as a function of squared lattice spacing $a^2$,
where the finer lattice results are on the measured point and the coarser
lattice results are obtained by linear-interpolation from the nearest two
points\footnote{
The matching here is done with a constant bare mass in units of MeV.
The logarithmic correction to an ideal matching with the renormalized mass
should be negligible for this qualitative study, 
given that the mass dependence of topological susceptibility is mild in
the region in question.
}.
The GL-DW result develops a large discretization error, and it
gets close to OV-OV towards the continuum limit. The OV-OV result is
more stable against lattice spacing. All results suggest $\chi_t$ is
vanishing in the continuum limit.

Focusing on the OV-OV result in the left panel the mass dependence of
the topological susceptibility indicates two regions for mass: one is
$0<m\lesssim 10$ MeV where the observation of continuum scaling above
strongly suggests $\chi_t=0$. Actually, $\chi_t$ with OV-OV is 
consistent with zero in this region. The other is $m\gtrsim 10$ MeV where
$\chi_t$ is significantly non-zero.
We note that the existence of the boundary at non-zero $m$ is also
suggested from 
GL-DW. While $\chi_t>0$ for $0<m\lesssim 10$ MeV, it is almost
constant. 
For $\chi_t\gtrsim 10$ MeV sudden development of $\chi_t$ is
observed. 
Due to its better precision over OV-OV, GL-DW results may be useful to
identify the location of the boundary.

We note that a preliminary computation of the pion mass on the zero temperature
configuration leads to an estimate of the physical $ud$ quark mass 
as $m=4$ MeV for the bare mass, which is well
inside the region where $\chi_t=0$ is suggested.

\begin{figure}[thb]
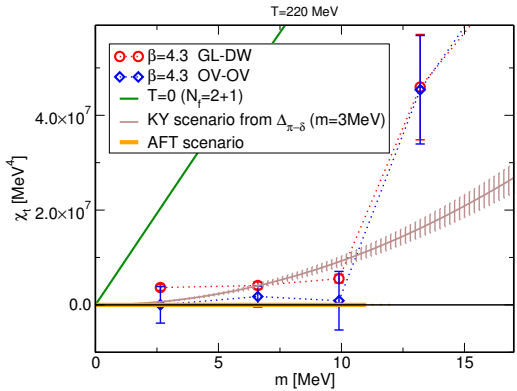

 \centering
 \sidecaption
 \includegraphics[width=0.475\textwidth,clip]{{{figures/chi-mf_beta4.3+T=0_mag2++}}}
 \caption{Topological susceptibility $\chi_t$ at $T\simeq 220$ MeV with
 possible scenarios based on Aoki-Fukaya-Taniguchi \cite{Aoki:2012yj}
 (orange) and Kanazawa-Yamamoto \cite{Kanazawa:2015xna} (brown). A
 zero-temperature result 
 \cite{Aoki:2017paw} for $N_f=2+1$ is plotted as a reference (green).}
 \label{fig:chit_scenario}
\end{figure}

Figure \ref{fig:chit_scenario} shows a magnified view of the left panel
of Fig.~\ref{fig:chit220} without GL-OV and OV-DW. The newly added green
line shows a zero temperature reference represented as a two-flavor ChPT
fit with $N_f=2+1$ results \cite{Aoki:2017paw}.
In this figure two scenarios are compared: one is 
Aoki-Fukaya-Taniguchi \cite{Aoki:2012yj} (AFT), where they claim that
the derivatives of $\chi_t$ with respect to quark mass vanish.
With $\chi_t=0$ at $m=0$ a natural solution would be $\chi_t=0$ for
$m < m_c$. The OV-OV result is consistent with this picture
with $10\lesssim m_c\lesssim 12$ MeV. 
The AFT result is based on the analyticity of Dirac eigenvalue spectral
density $\rho(\lambda)$.
On the other hand, Kanazawa-Yamamoto \cite{Kanazawa:2015xna} (KY) claims
that $U_A(1)$ should be violated for $T>T_c$ due to its violation in the
high enough temperature claimed in \cite{Laine:2003bd,Dunne:2010gd}.
They reported that the analyticity of $\rho(\lambda)$ needs to be
abandoned for the $U_A(1)$ violation. There is a KY scenario of $\chi_t(m)$
given in \cite{Kanazawa:2015xna}. 
To evaluate $\chi_t(m)$, one needs to
know the value of a low energy constant, which may be extracted from the
$U_A(1)$ order parameter measured with fixed topology.
At the lightest mass where the topological charge is practically fixed at
$|Q_t|=0$ after the reweighting (see Fig.~\ref{fig:hstg_OV_001}), we take
their proposal with our $U_A(1)$ breaking parameter 
$\Delta_{\pi-\delta}$ \cite{Lat2017Suzuki} and obtain the brown curve
($\propto m^2$) in the figure. This curve shows how $\chi_t(m)$ behaves
if the $U_A(1)$ symmetry were violated in the thermodynamic limit. 
Comparing with our OV-OV result, it has a tension ($>2\sigma$) at 
$m\simeq 13$ MeV.

\subsection{Topological susceptibility for  $T\gtrsim 220$ MeV}

\begin{figure}[thb]
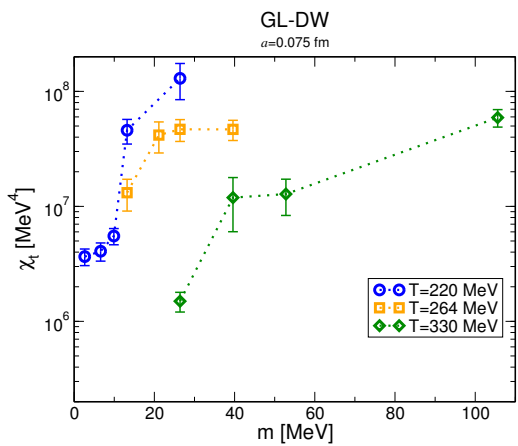

 \centering
 \sidecaption
 \includegraphics[width=0.475\textwidth,clip]{{{figures/chi_GLQ-DW-mf_beta4.3_3Ts_logy}}}
 \caption{Topological susceptibility $\chi_t$ at $T\simeq 220$, 264, 330
 is plotted as function of quark mass. Only results obtained with
 a gluonic operator without reweighting are shown. Gauge coupling is fixed
 and $a{-1}\simeq 2.64$ GeV for all.}
 \label{fig:chit_3Ts}
\end{figure}

To check whether the jump of the topological
susceptibility observed at $T\simeq 220$ MeV persists at other temperatures,
ensembles with different $N_t$ with fixed $\beta=4.3$ have been
generated and analyzed. The additional lattices are $N_t=10$ and 8 with
fixed $L=32$ as for $N_t=12$. The corresponding temperatures are
$T\simeq 264$ and 330 MeV respectively. Figure \ref{fig:chit_3Ts} shows the
topological susceptibility as a function of quark mass for three different
temperatures, where only GL-DW data are shown. 
A similar jump of $\chi_t$ at finite quark mass is
observed also for 
$T\simeq 264$ and 330 MeV. The position of the jump shifts
toward larger mass as $T$ is increased.

\section{Summary and outlook}\label{sec:summary}

Topological susceptibility $\chi_t$ in $N_f=2$ QCD was examined at temperatures
above the critical one with M\"obius domain wall fermion ensembles
reweighted to the overlap fermion ensembles. A special focus 
is put on the $T\simeq 220$ MeV ensembles with $N_t=12$. The preliminary results
suggest that for the range of bare mass $0\le m\lesssim 10$ MeV (which
includes physical $ud$ mass $m\simeq 4$ MeV) $\chi_t=0$ and for 
$m\gtrsim 10$ MeV a sudden development of $\chi_t$ starts.
It is consistent with the prediction of Aoki-Fukaya-Taniguchi
\cite{Aoki:2012yj} with
$U_A(1)$ symmetry restoration in the chiral limit, thus consistent with
the direct measurement of the order parameter of $U_A(1)$ \cite{Lat2017Suzuki}.
If that were due to finite volume effects and eventually we were to see
the breaking in the thermodynamic limit, then Kanazawa-Yamamoto
\cite{Aoki:2012yj} explains
how the $U_A(1)$ order parameter at finite volume is related to
$\chi_t$. The result has a $>2 \sigma$ tension.

We have examined the stability of the observation of the $\chi_t=0$ region 
with a comparison to the coarse lattice result at approximately
the same temperature, which indeed suggests the result is robust in the
continuum limit. However this comparison is done with the lattice site
number in the spatial direction fixed, therefore the physical box sizes
are different. We are now examining the volume dependence on the finer
lattice used in this report to check this.

Higher temperatures $T\simeq$ 264 and 330 MeV are also studied
with fixed lattice spacing $a^{-1}\simeq 2.64$ GeV.
A sudden change of $\chi_t$ as a function of the quark mass is also observed
for these temperatures. The point where the change occurs shifts
towards larger mass for higher temperature.
To get more insight for this observation, a systematic study for these
high temperatures in conjunction with the $U_A(1)$ order parameter is planned.

\section*{Acknowledgments}
We thank the members of the JLQCD collaboration for their support on
this work.
Numerical calculations are performed on the Blue Gene/Q at KEK under its
Large Scale Simulation Program (No. 16/17-14), Camphor 2 at the Institute
for Information Management and Communication, Kyoto University,
and Oakforest-PACS supercomputer operated by the Joint Center for Advanced High
Performance Computing (JCAHPC). This work is supported in part by JSPS
KAKENHI Grant Nos. JP26247043, 16K05320 and by the Post-K supercomputer
project through the Joint Institute for Computational Fundamental Science
(JICFuS).

\bibliography{all,lattice2017}

\end{document}